\documentclass[prd,showpacs,floatfix,superscriptaddress,twocolumn]{revtex4}
\usepackage{bm}
\usepackage{amsmath}
\usepackage{amssymb}
\usepackage{latexsym}
\usepackage{graphicx}

\newcommand{\dd}{\mathrm{d}}
\newcommand{\ee}{\mathrm{e}}

\begin{document}

\title{Emission of Low--Energy Photons by Electrons at Electron--Positron\\
and Electron-Ion Colliders with Dense Bunches}

\newcommand{\addrMST}{%
Department of Physics, Missouri University of Science and Technology,
Rolla, Missouri, MO65409, USA}

\newcommand{\addrHDphiltheo}{Institut f\"ur Theoretische Physik,
Universit\"{a}t Heidelberg,
Philosophenweg 16, 69120 Heidelberg, Germany}

\newcommand{\addrNSB}{Novosibirsk State University,
630090 Novosibirsk, Russia}

\author{U. D.~Jentschura}
\affiliation{\addrMST}
\affiliation{\addrHDphiltheo}
\author{G. L. Kotkin}
\affiliation{\addrNSB}
\author{V. G. Serbo}\email{serbo@math.nsc.ru}
\affiliation{\addrNSB}
\affiliation{\addrHDphiltheo}

\begin{abstract}
Usually, the emission of low-energy photons in electron-positron (or
electron-ion) bunch collisions is calculated with the same approach
as for synchrotron radiation (beamstrahlung). However, for soft
photons ($E_\gamma < E_{\rm c}$ where $E_{\rm c}$ is a critical
photon energy), when the coherence length of the radiation becomes
comparable to the bunch  length, the beamstrahlung approximation
becomes invalid. In this paper, we present results of our
calculation for this region based on approximation of classical
currents. We consider several colliders with dense bunches. The
number  of low-energy photons $\dd N_\gamma$ emitted by $N_e$
electrons per bunch crossing in the energy interval $\dd E_\gamma$
is $\dd N_\gamma = \alpha\, g \, N_e\, \dd E_\gamma/E_\gamma$, where
$\alpha$ is the fine-structure constant, and the function $g$, which
depends on the bunch parameters, typically is of order unity for
modern colliders. In particular, for the International Linear Collider
(ILC), we find that $E_{\rm c}= 83$~keV and $g= 5.5$ 
at a vanishing beam axis displacement, and
$g = 0.88$, $E_{\rm c}=0.24$ keV for KEKB. We also calculate the
specific dependence of $\dd{N}_\gamma$ on the impact parameter
between the two beam axes. In principle, the latter aspect allows
for online monitoring of the beam axis displacement.
\end{abstract}


\maketitle

%
%
\section{INTRODUCTION}

The importance of electromagnetic processes during bunch collisions at
colliders is well appreciated. Indeed, photons emitted in the collision of
dense bunches consisting of electron or positrons are a source for a number of
problems related to energy losses and background.  In particular, at the
International Linear Collider (ILC), we have to know the spectrum of photons
down to an energy of about 1~eV because such photons can produce high-energy
photons after Compton scattering.  As shown here,  for photon energies below
about $100$~keV, the usual beamstrahlung treatment of photoemission breaks down
because the coherence length becomes comparable to the bunch length.  A
treatment based on classical currents is used here in order to describe the
long-coherence-length limit of bremsstrahlung, where the electrons interact
with the collective field of the incoming positron (or ion) bunch.

In addition to a calculation of the spectrum of low energy photons, we find a
specific dependence of the photon rate on the impact parameters between the two
bunch axes, namely, we find that the number of emitted photons exhibits a
maximum at a non-vanishing beam axis displacement, even though the luminosity
is a monotonically decreasing function of the shift of the beam axes.  This
anomaly finds a natural explanation in a classical treatment.

Here, in addition to the ILC, we consider several colliders
which have dense bunches of particles: the existing
$e^+e^-$ storage rings KEKB, BEPC, VEPP-2000 and the
currently envisaged $e$-Au and $e$-$p$ beam collision
options at the RHIC collider (collider parameters
as taken from Refs.~\cite{ILC,YaEtAl2006,eRHIC}
are listed in Table~\ref{Table1}).
\begin{table*}[htb]
\begin{center}
\begin{minipage}{12cm}
\begin{center}
\caption{\label{Table1} Parameters of the discussed
colliders. We use the following notations: $E_e$ ($E_p$)
and $N_{e}$ ($N_{p}$) are the energy and the number of
particles in the electron bunch (subscript~$e$)
and the positively charged bunch (subscript~$p$, positrons or ions).
We denote by $\sigma_{z}$ the longitudinal, and by
$\sigma_x$ and $\sigma_y$ the horizontal and vertical
transverse sizes of the bunch.}
\renewcommand{\arraystretch}{1.5}
\begin{tabular}{c@{\hspace{0.3cm}}c@{\hspace{0.3cm}}c@{\hspace{0.3cm}}%
c@{\hspace{0.3cm}}c@{\hspace{0.3cm}}c}
\hline
\hline
Collider &  $E_e/E_p\;$ (GeV) & $N_e/N_p\;(10^{10})$ & $\sigma_z $ (cm) &
$\sigma_{x}\;$ ($\mu$m) & $\sigma_{y}\;$ ($\mu$m)
\\
\hline
ILC $e^-/e^+$  & $250/250$   & $2/2$ & $0.3$  & $0.64$ & $0.0057$ \\
KEKB $e^-/e^+$ & $8/3.5$ & $6/8$ & $0.65$ & $110$  & $2.4$ \\
BEPC $e^-/e^+$ & $2/2$     & $13/13$  & $5$    & $890$  & $37$ \\
VEPP-2000 $e^-/e^+$ & $1/1$ & $16/16$ & $4$ & $125$ & $125$ \\
eRHIC $e/{\mathrm{Au}}$ & $10/100\,A$ & $10/0.1$ & $20$ & $50$ & $50$ \\
eRHIC $e/p$ & $10/250$ & $10/10$ & $10$ & $40$ & $40$ \\
\hline
\hline
\end{tabular}
\end{center}
\end{minipage}
\end{center}
\end{table*}

\begin{table}[htb]
\caption{\label{Table2} Characteristics of low-energy
photon emission for the discussed colliders.
The beam aspect ratio
$\varepsilon$ is defined in Eq.~(\ref{ar}) below.}
\renewcommand{\arraystretch}{1.5}
\begin{center}
\par
\begin{tabular}{c@{\hspace{0.3cm}}c@{\hspace{0.3cm}}c@{\hspace{0.3cm}}c}
\hline
\hline
Collider &  $\eta$ & $E_{\rm c}\;$ (keV) & $g(\eta, \varepsilon)$ \\
\hline
ILC  & $87$ & $83$ & $5.5$ \\
KEKB & 2.0 & $6.0$ & $0.88$ \\
BEPC & $0.61$ & $0.24$ & $0.16$ \\
VEPP-2000 & $1.8$ & $0.018$ & $0.78$ \\
eRHIC $e/{\mathrm{Au}}$ & $2.7$ & $4.0$ & $1.2$ \\
eRHIC $e/p$ & $3.3$ & $2.8$ & $1.4$ \\
\hline
\hline
\end{tabular}
\end{center}
\end{table}

For definiteness, we start with the ILC collider. An
important process at this collider is the emission of
low-energy photons by electrons in the collective electromagnetic
field of a positron bunch (see, for example,
Ref.~\cite{ZoKuSe1981}). Certainly, in this case the emission of
photons by positrons is of course equally important, but we
here restrict our attention to the emission of photons
by the electrons. These photons with an energy $E_\gamma$
down to $1$~eV may lead to Compton scattering with
particles of the incoming bunch, resulting in the
production of photons with the energy $E_\gamma \sim 100$
GeV via the Compton process. The corresponding Compton
cross section is quite large, of the order of $10^{-25}$~cm$^2$.
Thus, the calculation of the low-energy
photon rate is important.

Usually, the emission of photons at linear colliders is
calculated using the same approach as for synchrotron
radiation. In this case, the emission is called
beamstrahlung (see Ref.~\cite{Ch1988}). This approximation is
good for photons in an energy range from a few MeV to
several tens of~GeV. However, at a smaller photon energy,
\begin{equation}
E_\gamma < E_{\rm c} \sim 100\,{\mathrm{keV}} \,,
\end{equation}
this approximation is invalid. Let us discuss this problem
in more detail.

The electric ${\bm E}$ and magnetic ${\bm B}$ fields of the
positron bunch are approximately equal in magnitude
(see also Table~\ref{Table1} for numerical data),
\begin{equation}
|{\bm E}| \approx |{\bm B}| \sim \frac{e \, N_p}%
{(\sigma_x+\sigma_y) \, \sigma_z}\,.
\label{EB}
\end{equation}
These fields are transverse, and they deflect the electron
into the same direction. In such fields, the electron moves
around a circumference of radius $\rho\sim \gamma_e \, m_e
\, c^2/(eB)$ in the sense of a relativistic ``cyclotron''
motion, where $\gamma_e = E_e/(m_e \, c^2)$ is the Lorentz
factor. The electron gets a deflection angle $\theta_e \sim
{\sigma_z}/{\rho}$ or
\begin{equation}
\theta_e \sim \frac{\eta}{\gamma_e},
\label{1}
\end{equation}
where the important parameter
\begin{equation}
\eta=\frac{r_e \, N_p}{\sigma_x+\sigma_y}
\label{defeta}
\end{equation}
is introduced. Here, $r_e = e^2/(m_e c^2)$ is the classical
electron radius. The dimensionless parameter $\eta$
measures the ``thickness'' of the positron bunch with
respect to the bending effect for the
classical electron trajectory. In the case of a
collision of electrons with ions of charge number $Z$,
the parameter $\eta$ acquires an additional factor to
read
\begin{equation}
\eta=\frac{Z \, r_e \, N_p}{\sigma_x+\sigma_y}\,.
 \label{defetaZ}
\end{equation}

Let us consider the emission of photons with small energy
$E_\gamma$ in an angular interval up to $\theta_e$. It is well
known (see Ref.~\cite{BeLiPi1982vol4}, \S~93) that the corresponding
coherence length is of the order of
\begin{equation}
\label{3a}
l_{\rm coh}(E_\gamma) \sim 
\frac{\hbar c}{(1-v_e \cos{\theta_e} / c)\,E_\gamma}\,,
\end{equation}
where $v_e$ is the velocity of the electron. Taking into account
that the electron Lorentz factor 
$\gamma_e = 1/\sqrt{1-(v_e/c)^2}\gg 1$
and that $\theta_e\ll 1$, we obtain
\begin{equation}
l_{\rm coh}(E_\gamma) \sim 
\frac{2 \gamma_e^2}{1+\gamma_e^2 \,
\theta_e^2} \, \frac{\hbar c}{E_\gamma} \, . 
\label{3}
\end{equation}
The critical photon energy $E_{\rm c}$ corresponds to the
case when the coherence length is of the order of half the
positron bunch length,
\begin{equation}
\label{crit}
l_{\rm coh}(E_{\rm c})\sim \tfrac12 \, \sigma_z\,,
\end{equation}
which leads to
\begin{equation}
E_{\rm c} =
\frac{4\gamma_e^2}{1+\eta^2}\; \frac{\hbar c}{\sigma_{z}}\,.
\label{Ec}
\end{equation}
When the coherence length is even larger than the positron bunch
length, then the beamstrahlung treatment becomes invalid, and the
spectrum of emitted photons is determined by coherent radiation from
the electron trajectory in the whole incoming bunch. By contrast,
the beamstrahlung approach is valid when the coherence length is
small in comparison to the bunch length.

If the parameter $\eta$ is small, $\eta\ll 1$, the number
of photons $dN_\gamma$ emitted per one bunch crossing is
proportional to the number of electrons and to the square
of the number of positrons:
\begin{equation}
\dd N_{\gamma}\; \propto \;N_e\, {N_p}^2\,
\frac{\dd E_{\gamma}}{E_{\gamma}} \, .
\label{5}
\end{equation}
In some respects, the radiation in this case is similar to
ordinary bremsstrahlung, therefore we call it coherent
bremsstrahlung (CBS). It differs substantially from
beamstrahlung. CBS was considered in detail in
Refs.~\cite{GiKoPoSe1992prl,GiKoPoSe1992plb,GiKoPoSe1993,GiKoPoSe1992yf1,GiKoPoSe1992yf2,KoSe2004}.

However, for all the discussed colliders, the parameter
$\eta \gtrsim 1$ (see Table~\ref{Table2}). In particular,
for the ILC collider the parameter $\eta$ is indeed large,
\begin{equation}
\eta = 87 \,,
\label{6}
\end{equation}
and the critical energy is
\begin{equation}
E_{\rm c}= 83 \, \mbox{keV}\,.
\label{7}
\end{equation}
Therefore, at a photon energy $E_\gamma < E_{\rm c}$, one
cannot use the formulae related to beamstrahlung or CBS. In
Ref.~\cite{ShTy2003}, it has been pointed out that an adequate
approach to the calculation of photoemission in this
energy range can be based on classical currents.
It means that one can use results for
the classical low-frequency emission taking into account
that the electron deflection angle $\theta_e$
for this process typically is large
in comparison with $1/\gamma_e$, but small in comparison
with unity:
\begin{equation}
\frac{1}{\gamma_e}\lesssim \theta_e \sim \frac{\eta}{\gamma_e}\ll 1
\label{8}
\end{equation}
(see Appendix~\ref{appa} for details).

For definiteness, we consider the emission of photons by
electrons and assume that the electron and positron bunches
perform a head-on collision along the $z$-axis. We also
assume that the transverse distribution of the particles in
the positron bunch does not change during the collision
(therefore, the impact parameter of the particles in the
bunch, here denoted as $\mbox{\boldmath $\varrho$}$, does
not change either). In numerical calculations, we assume
that the bunches in the interaction region have a Gaussian
particle distribution with a transverse density
$n_i({\mbox{\boldmath $\varrho$}})$ of the form
\begin{align}
n_i({\mbox{\boldmath $\varrho$}}) =& \;
\int_{-\infty}^{+\infty}n_i({\mbox{\boldmath
$\varrho$}},z)\,\dd z
\nonumber\\[2ex]
=& \; \frac{N_i}{2\pi \sigma_x \sigma_y}\, \exp\left[
-\frac{\varrho_x^2}{2\sigma_x^2} -
\frac{\varrho_y^2}{2\sigma_y^2}\right] \,,
\label{9}
\end{align}
with $i=e,\;p$. This implies that we neglect potential
angular dispersion in the beams and the pinch effect.

Our paper is organized as follows. In Sec.~\ref{sec2}, we calculate
the spectrum of low-energy photons and in Sec.~\ref{sec3}, we
consider the dependence of the photon rate on beam axis
displacement. Finally, some conclusions are drawn in
Sec.~\ref{sec4}. Throughout the text, we use Gaussian units with
the fine-structure constant $\alpha=e^2/(\hbar c)\approx
1/137$.

%
%
\section{SPECTRUM OF LOW-ENERGY PHOTONS}
\label{sec2}

In this section we consider beams at zero displacement of
the beam axes $R = |{\bm R}| = 0$. In that setting, let an
electron with the relative impact parameter
$\mbox{\boldmath $\varrho$}$ with respect to the beam axis
deflect in the electromagnetic field of the positron bunch
at an angle $\theta_e$ given by Eq.~(\ref{8}) and acquire a transverse
momentum ${\bm p}_\perp$. The probability of emitting a
low-energy photon $\dd w$ depends on the dimensionless
vector parameter
\begin{equation}
\mbox{\boldmath $\xi$}(\mbox{\boldmath $\varrho$}) =
\frac{{\bm p}_\perp}{2 \, m_e \, c} =
-\,r_e\, \int \frac{{\mbox{\boldmath $\varrho$}} -
{\mbox{\boldmath $\varrho$}}'}%
{({\mbox{\boldmath $\varrho$}} -
{\mbox{\boldmath $\varrho$}}')^2}
\, n_p({\mbox{\boldmath $\varrho$}}')\, \dd^2 \varrho'
\label{xi}
\end{equation}
and is equal to (see Appendix~\ref{appa})
\begin{equation}
\dd w = \alpha\,G(\xi) \, \frac{\dd E_\gamma}{E_\gamma}\,,
\end{equation}
where $\xi \equiv | \bm{\xi}(\bm{\varrho}) |$, and the
function $G(\xi)$ is given by
\begin{equation}
G(\xi) = \frac{2}{\pi} \, \left[ \frac{2 \, \xi^2 + 1}{\xi
\,\sqrt{\xi^2+1}}\, \ln\left(\xi + \sqrt{\xi^2+1}\right) -1
\right]\,. \label{Fxi}
\end{equation}

The number of low-energy photons, produced by the electron
bunch, is obtained by an integration over the distribution
of particles in the electron bunch. As a result, we
find a simple expression,
\begin{equation}
\dd N_\gamma = \alpha\, g\, N_e \, \frac{\dd
E_\gamma}{E_\gamma}\,, \label{defg}
\end{equation}
where all nontrivial information about the
bunches is accumulated in the function
\begin{equation}
g = \langle G(\xi)\rangle= \int G(\xi)\,
\frac{n_e({\mbox{\boldmath $\varrho$}})}{N_e}\, \dd^2
\varrho\,. \label{dNgamma2}
\end{equation}
For identical Gaussian beams and $R =0$, the function $g$
only depends on the dimensionless parameter $\eta$, defined
in Eq.~(\ref{defeta}), and the beam aspect ratio
\begin{equation}
\varepsilon = \frac{\sigma_y}{\sigma_x}\leq 1\,,
 \label{ar}
\end{equation}
i.e.
\begin{equation}
g=g(\eta,\,\varepsilon)\,.
 \label{gee}
\end{equation}
Calculated values of $g(\eta,\,\varepsilon)$ are given in
Table~\ref{Table2} for various colliders. Details of the
calculation can be found in Appendix~\ref{appb}.

\begin{figure}[htb]
\includegraphics[width=1.0\linewidth]{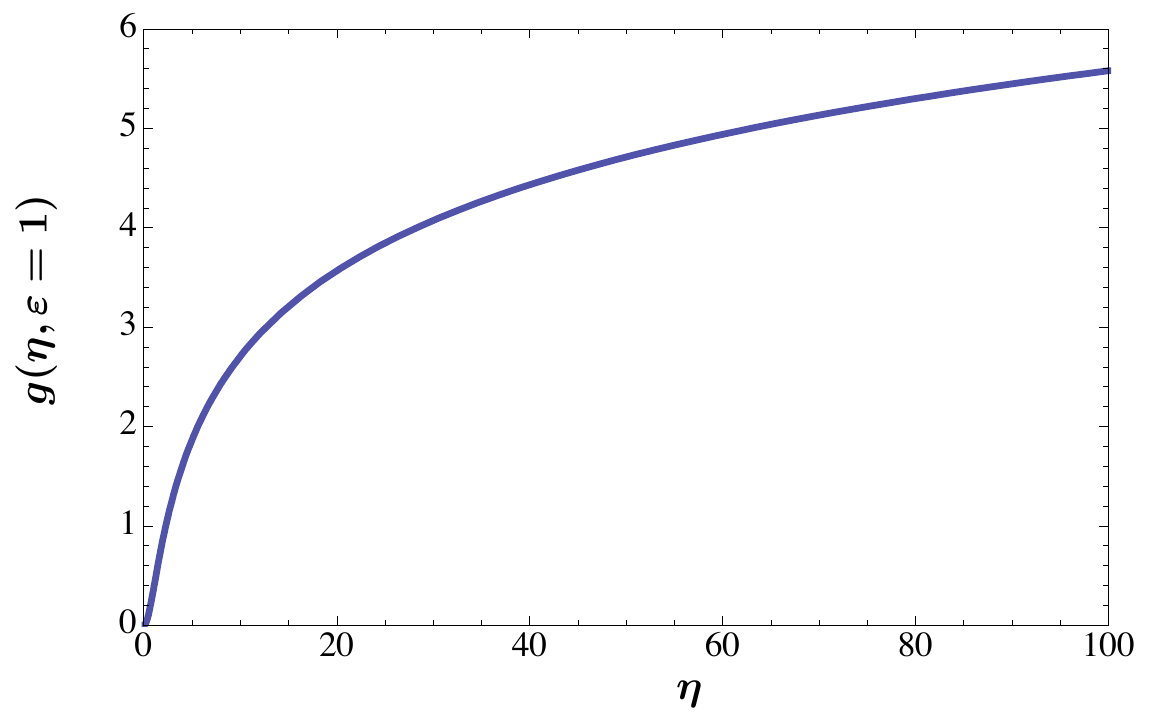}
\caption{\label{Fig1} (Color online) The function
$g(\eta,\,\varepsilon=1)$ for round identical beams.}
\end{figure}

For round beams, $\sigma_x = \sigma_y$, the dependence of
$g$ on $\eta$ is shown on Fig.~\ref{Fig1}. We checked by
numerical calculations that the dependence of this function on
the aspect ratio $\varepsilon$ is very weak: when
$\varepsilon$ varies from $1$ to $0.01$, the function
$g(\eta,\varepsilon)$ changes only by less than one
percent for $1 < \eta < 100$, which is the
relevant parameter range for all colliders
listed in Table~\ref{Table2}. It
means that Fig.~\ref{Fig1} gives a very good approximation
to the dependence of $g$ on
$\eta$ not only for round but for flat beams as well.

For small parameters $\eta\ll 1$, the $g$ function varies
quadratically with the number of particles in the positron
bunch [see Eq.~(\ref{smalleta}) in Appendix~\ref{appb}],
\begin{equation}
g(\eta,\,\varepsilon) = \frac{8}{3\pi}\langle \xi^2\rangle
\approx 0.5\,\eta^2 \propto N_p^2\,.
 \label{g-small-eta}
\end{equation}
The region where the above quadratic approximation is valid
is confined to the region $\eta \lesssim 1$ and thus not
visually discernible in Fig.~\ref{Fig1}. This result is
fully consistent with that obtained in the CBS
approximation of Refs.~\cite{GiKoPoSe1992prl,GiKoPoSe1992plb,GiKoPoSe1993,GiKoPoSe1992yf1,GiKoPoSe1992yf2}, because in the
limit $\eta \to 0$, mainly small deflection angles
contribute to the coherent process, and thus the CBS
approximation remains valid, including prefactors. For
large parameters $\eta\gg 1$, by contrast, a logarithmic
dependence on $\eta$ is found [see Eq.~(\ref{largeeta}) in
Appendix~\ref{appb}],
\begin{equation}
g(\eta, \,\varepsilon) \approx \frac{2}{\pi}\left[ \langle
\ln{(4\xi^2)} \rangle - 1 \right] \approx
\frac{2}{\pi}\left[ \ln(\eta^2) - 0.34 \right] \,.
\label{g-large-eta}
\end{equation}
In this case, in view of Eq.~(\ref{defeta}), the dependence
on $N_p$ for large $N_p$ is also logarithmic and thus much
weaker than previously,
\begin{equation}
g(\eta, \, \varepsilon) \propto \ln{\left(N_p^2\right)}\,.
\label{g-large-eta-Np}
\end{equation}
In that sense, the coherent effects should be less
pronounced for the ILC as compared to KEKB.

\begin{figure}[htb]
\centering
\includegraphics[angle=0,width=1.0\linewidth]{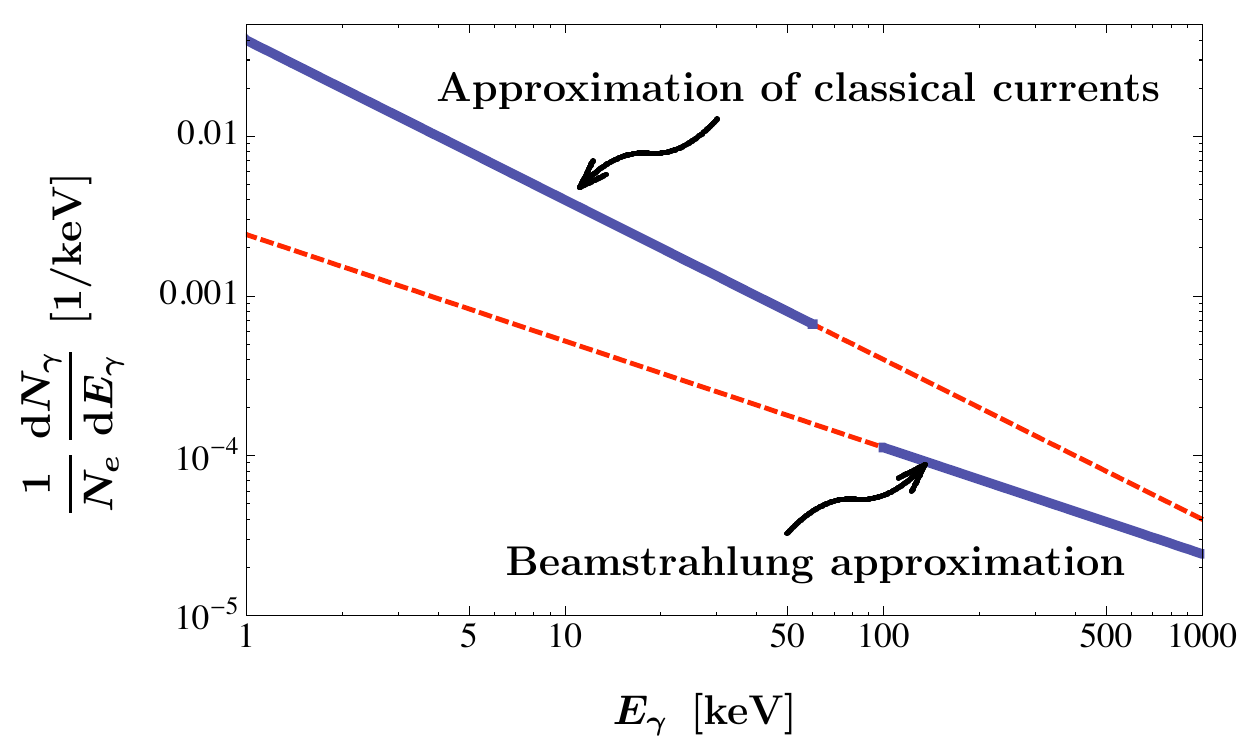}
\caption{\label{Fig2} (Color online)
Spectrum of photons at the ILC
collider, produced by one electron, in the approximation of
classical currents [upper curve according to
Eq.~(\ref{defg})] and in the beamstrahlung approximation
[lower curve according to Eq.~(\ref{dNgammaBS})]. In the
photon energy domains where the approximations are
justified, the curves are drawn as solid lines. In those
photon energy regions where the approximations are not
applicable, the curves are drawn as dashed lines.}
\end{figure}

In particular, we have calculated that for ILC (see
Table~\ref{Table2}) $g = 5.5$. Let us compare this result
with the corresponding one in the beamstrahlung case. A
rough estimation of the beamstrahlung rate can be performed
using known formulae for synchrotron radiation (SR). A
single electron, moving through a uniform external magnetic
field $B_{\rm ext}$, emits $\dd n_\gamma^{\rm SR}$ photons
in the time interval $\Delta t$ (see Sec.~14.6 of
Ref.~\cite{Ja1998}):
\begin{align}
\dd n_\gamma^{\rm SR} =& \; \frac{\dd I}{E_\gamma} \,
\Delta t
\nonumber\\[2ex]
=& \; \frac49 \, \alpha \, F\left({E_\gamma}/{E_{\rm
c}^{\rm SR}} \right) \frac{\dd E_\gamma}{E_\gamma}\,
\frac{e \, B_{\rm ext}}{m_e \, c}\, \Delta t \, .
\label{121}
\end{align}
Here, $\dd I$ is the classical synchrotron radiation
intensity, and $E_{\rm c}^{\rm SR}$ is the critical
synchrotron energy
\begin{equation}
E_{\rm c}^{\rm SR} =  \frac 32 \,
\gamma_e^2 \,
\left( \frac{\hbar}{m_e \, c} \right) \, e \, B_{\rm ext} \,.
\label{122}
\end{equation}
The normalized function $F(y)$ is defined as an
integral over the modified Bessel function $K_{5/3}(x)$,
\begin{align}
F(y) =& \;  \frac{9 \, \sqrt{3}}{8 \pi} \, y \,
\int_y^{\infty} K_{5/3}(x) \, \dd x
\nonumber\\[2ex]
\approx & \; 1.33\, y^{1/3} \quad \mbox{for} \quad y \ll 1 \ .
\label{123}
\end{align}
To estimate the number of low-energy photons produced by
the electron bunch moving through the positron bunch, we
use the mean value of the critical beamstrahlung energy
(see Ref.~\cite{Ch1988}),
\begin{align}
E_{\rm c}^{\rm BS} =& \;
\frac54 \, \gamma_e^2 \,
\left( \frac{\hbar}{m_e c} \right)
\left( \frac{e^2 \, N_p}{\sigma_z \, (\sigma_x+\sigma_y)} \right)
\nonumber\\[2ex]
=& \; \frac{5}{16} \, \eta \, (1 + \eta^2) \, E_{\rm c}\,.
 \label{EcBC}
\end{align}
Here, we have used the relations (\ref{EB}) and (\ref{Ec}),
and the flight time through the beam $\Delta t \sim
{\sigma_z}/{c}$. As a result, we obtain an estimate for the
number of beamstrahlung photons emitted per bunch
collision,
\begin{eqnarray}
\dd N_\gamma^{BS} &\approx& \alpha\, \eta\,
\left(\frac{E_\gamma}{E_{\rm c}^{\rm BS}} \right)^{1/3}\,
N_e\,\frac{\dd E_\gamma}{E_\gamma} \nonumber
\\
&\approx& 1.5 \, \alpha\, \left(\frac{E_\gamma}{E_{\rm
c}}\right)^{1/3}\, N_e\,\frac{\dd E_\gamma}{E_\gamma}
\label{dNgammaBS}
\end{eqnarray}
with $E_{\rm c}^{\rm BS}=18$~GeV for ILC, which is much higher than
the entry $E_{\rm c} =83$~keV from Table~\ref{Table2}. The power-law
behavior of the results is clearly visible on a logarithmic scale as
plotted in Fig.~\ref{Fig2}. We have compared our approximate,
analytic results for beamstrahlung to those of elaborate
Monte--Carlo simulations of V.~Telnov~\cite{Te1995}, designed for
the modeling of linear colliders, and we may state here that
reassuringly, we have found good agreement on level of 10--20
percent.

For a photon energy less than $E_{\rm c}=83$~keV, the ratio $\dd
N_\gamma^{BS}/\dd N_\gamma$ from Eq.~(\ref{dNgammaBS}) and
Eq.~(\ref{defg}) is considerable less then unity (see
Fig.~\ref{Fig2}), because of the smallness of the factor
$1.5\,(E_\gamma/E_{\rm c})^{1/3}$ as compared with $g=5.5$.

We would also like to remark that the results in
Eq.~(\ref{dNgammaBS}) have the ``wrong'' low-energy asymptotics (the
number of radiated photons is finite in the infrared limit), whereas
the classical-current result (\ref{defg}) restores the well-known
infrared ``catastrophe,'' as it should.

%
%
\section{BEAM AXIS DISPLACEMENT}
\label{sec3}

Let us now consider the case, when the electron bunch axis
is shifted by a distance ${\bm R} = (R_x, R_y)$
with respect to the
positron bunch axis. In this case, the transverse density
of the electron bunch has the form
\begin{eqnarray}
\label{dis}
n_e({\mbox{\boldmath $\varrho$}})&=&
\frac{N_e}{2\pi \sigma_{x} \sigma_{y}}
 \\
&\times&\exp\left[ -
\frac{(\varrho_x-R_x)^2}{2\sigma_{x}^2}-
\frac{(\varrho_y-R_y)^2}{2\sigma_{y}^2}\right]\,.
 \nonumber
\end{eqnarray}

The luminosity of the beams is determined by the expression
\begin{equation}
L({\bm R})\ = \nu \,\int n_e(\mbox{\boldmath
$\varrho$})\,n_p(\mbox{\boldmath $\varrho$})\,\dd^2 \varrho\,,
 \label{lum}
\end{equation}
where $\nu$ is the collision rate. For identical Gaussian
beams, the luminosity reads
\begin{subequations}
\begin{eqnarray}
L({\bm R})\; &=& \; L(0) \, \exp\left( -\frac{R_x^2}{4\sigma^2_x} -
\frac{R_y^2}{4\sigma^2_y} \right)\,, \label{lumindec}
\\
L(0)&=&\nu \frac{N_e N_p}{4\pi \sigma_{x} \sigma_{y}} \,.
\end{eqnarray}
\end{subequations}
Here, $L(0)$ is the luminosity without displacement. The
important scaling function $g$ was defined in
Eq.~(\ref{defg}) and was found to depend, for $R = 0$ (see
Sec.~\ref{sec2}), on $\eta$ and only slightly on the beam
aspect ratio $\varepsilon$ [see Eq.~(\ref{ar})]. Now, $g$
becomes a function of two more variables, namely the beam
axis displacement in the $x$ and in the $y$ direction, to
read
\begin{equation}
 g = g(\eta,\varepsilon, R_x, R_y)\,.
 \label{g-function}
\end{equation}
As discussed below, the dependence on $\varepsilon$ may
be quite important for non-vanishing values of ${\bm R}$,
in contrast to the case $R = 0$.

From Eq.~(\ref{lumindec}) we clearly see the following. If
the electron bunch axis is shifted by a distance $R$, the
luminosity decreases very quickly (exponentially). This
entails a corresponding decrease in the number of events
for ordinary reactions since this number is directly
proportional to the luminosity. However,
the emission of photons
(and especially low-energy photons!) is an extraordinary
reaction in that sense. The long-range Coulomb fields may
result in a more complicated relation between rate of photons
and the luminosity, namely, the rate of photons can decrease
more slowly than the luminosity or even increase with
$R$ in a certain range. This phenomenon has been
studied experimentally for ordinary incoherent
bremsstrahlung at the VEPP-4 collider~\cite{BlEtAl1982} and for
beamstrahlung at the SLC collider~\cite{BoEtAl1989}. A detailed
theoretical treatment of this effect for CBS was given in
Refs.~\cite{GiKoPoSe1992prl,GiKoPoSe1992plb,GiKoPoSe1993,GiKoPoSe1992yf1,GiKoPoSe1992yf2}. Therefore, in our case we also
expect an unusual behavior of the photon rate as a function
of the beam axis displacements.

%
%
\subsection{Flat beams: the ILC and KEKB colliders}
\label{flatbeams}

\begin{figure}[htb]
\centering
\includegraphics[angle=0,width=1.0\linewidth]{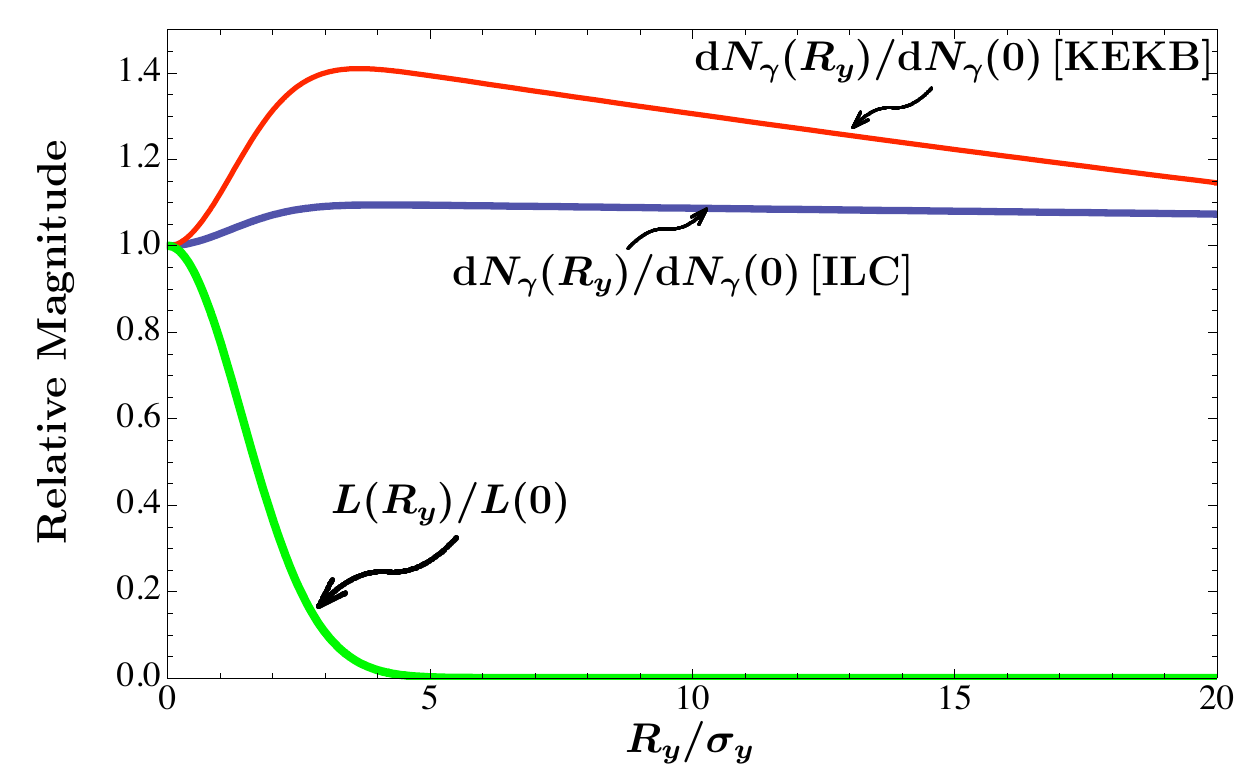}
\caption{\label{Fig3} (Color online)
Relative magnitude of luminosity and rate of
emitted photons as a function of the vertical beam axis
displacement $R_y$, for the ILC and KEKB colliders. The
two upper curves display the
normalized photon emission rate $\dd N_\gamma(R)/\dd
N_\gamma(0)=g(\eta,\varepsilon,
R_x=0,R_y)/g(\eta,\varepsilon, 0,0)$ for KEKB and ILC;
the lower curve shows the normalized
luminosity $L(R_y)/L(0)$.}
\end{figure}

\begin{figure}[htb]
\centering
\includegraphics[angle=0,width=1.0\linewidth]{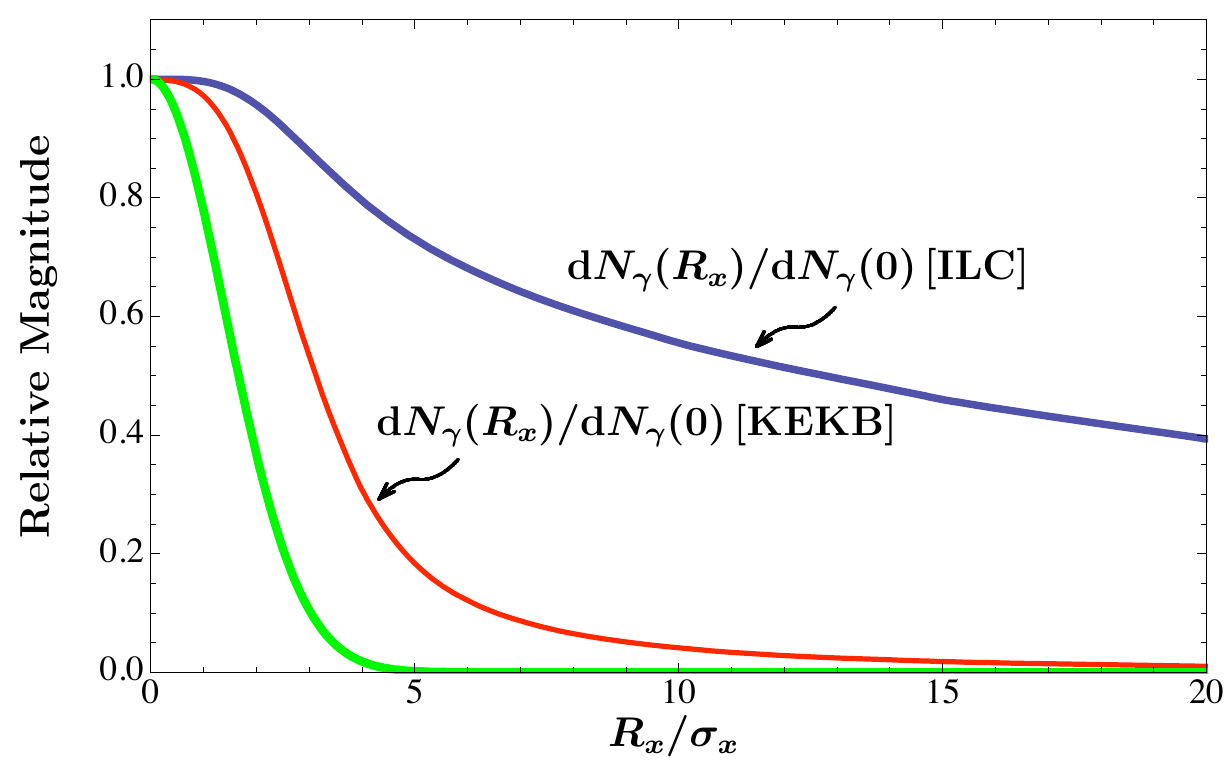}
\caption{\label{Fig4} (Color online)
The same as in Fig.~\ref{Fig3}, but for the
horizontal beam axis displacement $R_x$.
The unlabeled curve displays the relative luminosity
$L(R_x)/L(0)$ and has the same form as the
corresponding curve for $L(R_y)/L(0)$ in Fig.~\ref{Fig3}.}
\end{figure}

The electron and positron beams at the ILC and KEKB colliders
are flat, with their vertical sizes much smaller than the
horizontal ones (see also Table~\ref{Table1}),
\begin{equation}
\sigma_y \sim \frac{\sigma_x}{100}\,.
\end{equation}
We have calculated the photon rate at these colliders and
find a different behavior as a function of the
vertical and horizontal
displacements of the beam axes (see Fig.~\ref{Fig3} and
Fig.~\ref{Fig4}, respectively).

If the electron bunch axis is shifted in the vertical
direction by a distance $R_y$, the luminosity (and the
number of events for ordinary reactions) decreases very
quickly [see Eq. (\ref{lumindec})]:
\begin{equation}
L({\bm R})\; = \; L(0) \, \exp\left(
 - \frac{R_y^2}{4\sigma^2_y}
\right)\,.
\end{equation}
On the contrary, in the case of a vertical displacement,
the number of low-energy photons increases by about $9$ \%
at $R_{y}= 4\,\sigma _{y}$ for the ILC collider and by $40$
\% at $R_{y}= 3.5\,\sigma _{y}$ for the KEKB collider.
After that, the rate of photons decreases, but even at
$R_y=20\, \sigma_y$ it is still larger than at $R_y=0$ for
the ILC and for KEKB (by $7\,\%$ and $15\,\%$,
respectively). The corresponding curves are presented in
Fig.~\ref{Fig3}. If the electron bunch  axis  is  shifted
in the horizontal direction by a distance $R_x$, the photon
rate decreases, but more slowly as compared to the
exponential decrease of the luminosity. The corresponding
curves are presented in Fig.~\ref{Fig4}.

\begin{figure}[htb]
\centering
\includegraphics[angle=0,width=1.0\linewidth]{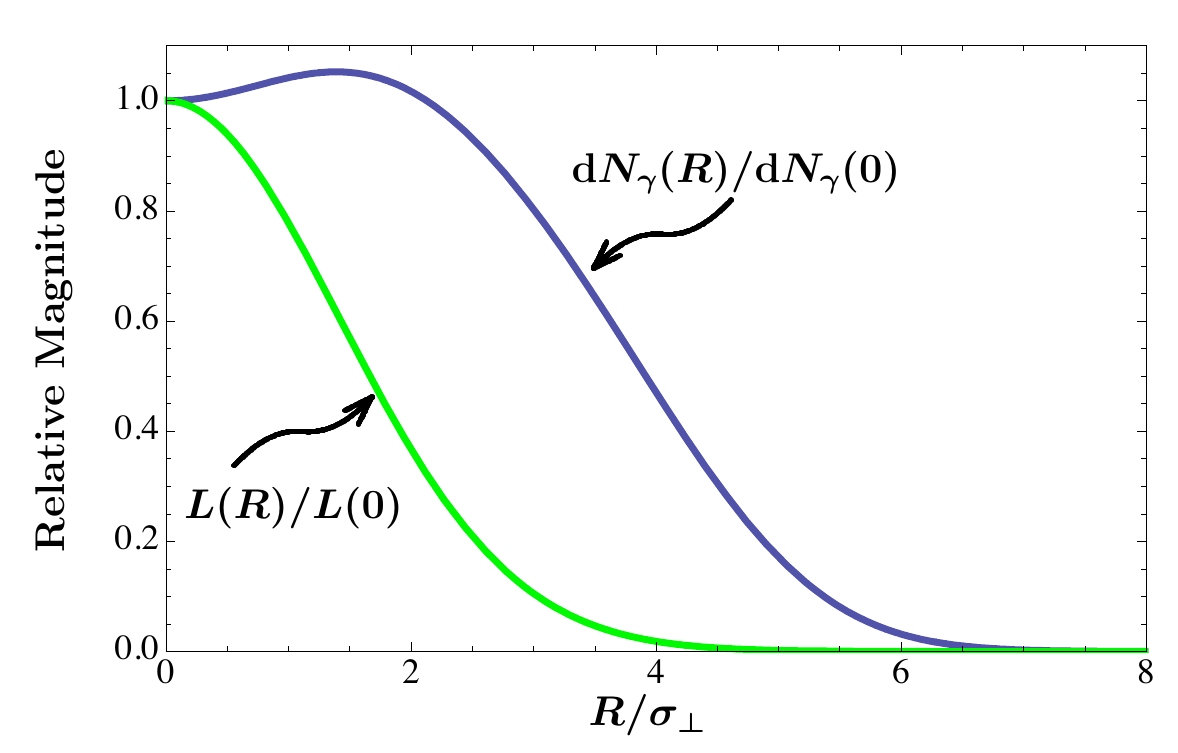}
\caption{(Color online) Relative magnitude of the luminosity and rate of
emitted photons as a function of the beam axis displacement
$R = \sqrt{R_x^2+R_y^2}$, for the VEPP-2000 collider. The
upper curve displays the normalized photon emission rate
$\dd N_\gamma(R)/\dd N_\gamma(0)=g(\eta,\varepsilon=1,
R_x,R_y)/g(\eta,\varepsilon=1,0,0)$ ; the lower curve shows
the normalized luminosity $L(R)/L(0)$. }
 \label{Fig5}
\end{figure}

\begin{figure*}[htb]
\begin{center}
\begin{minipage}[t]{0.49\linewidth}
\begin{center}
\includegraphics[width=1.0\linewidth]{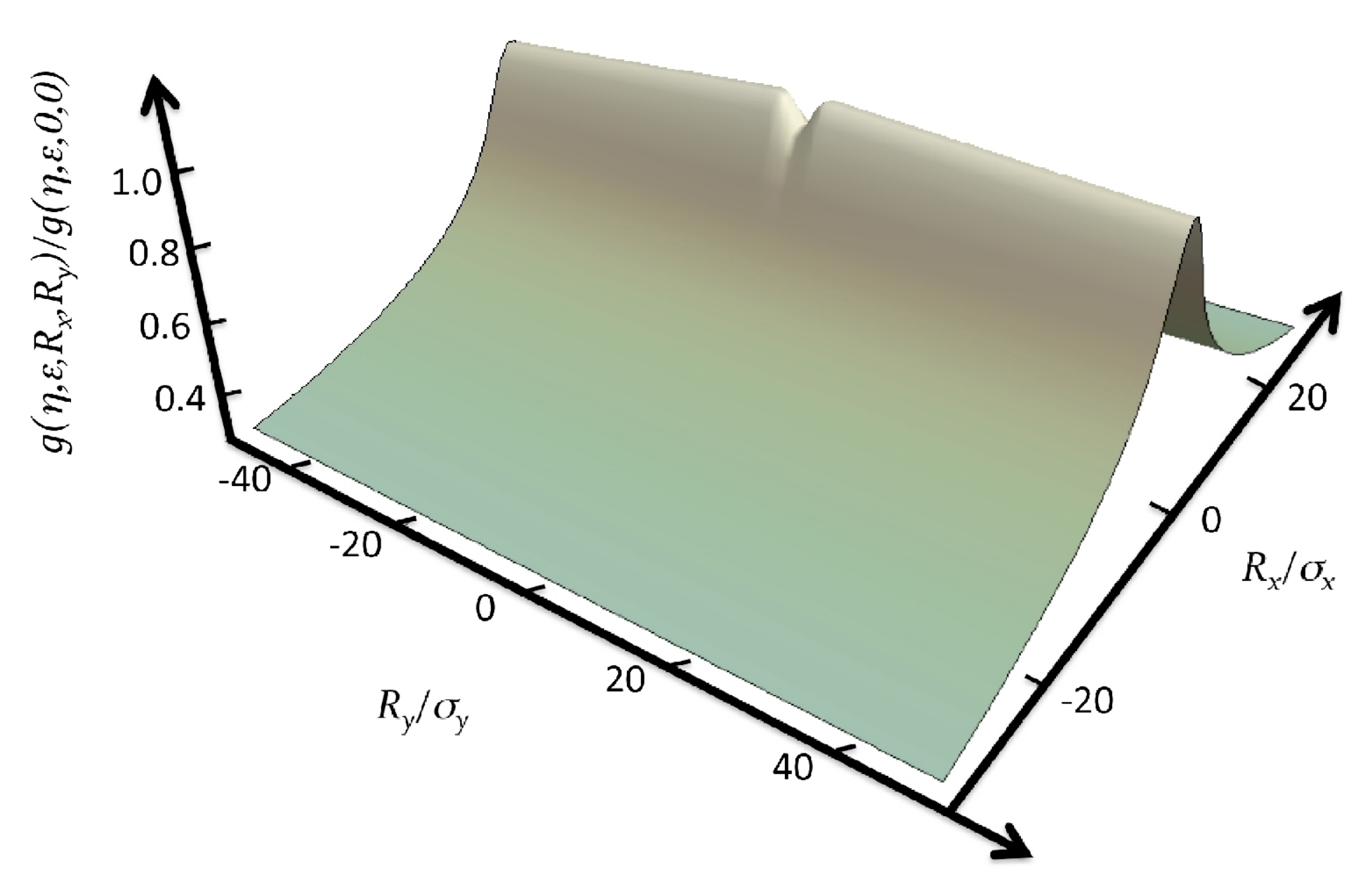} \\
{\bf (a)}
\end{center}
\end{minipage}
\begin{minipage}[t]{0.49\linewidth}
\begin{center}
\includegraphics[width=1.0\linewidth]{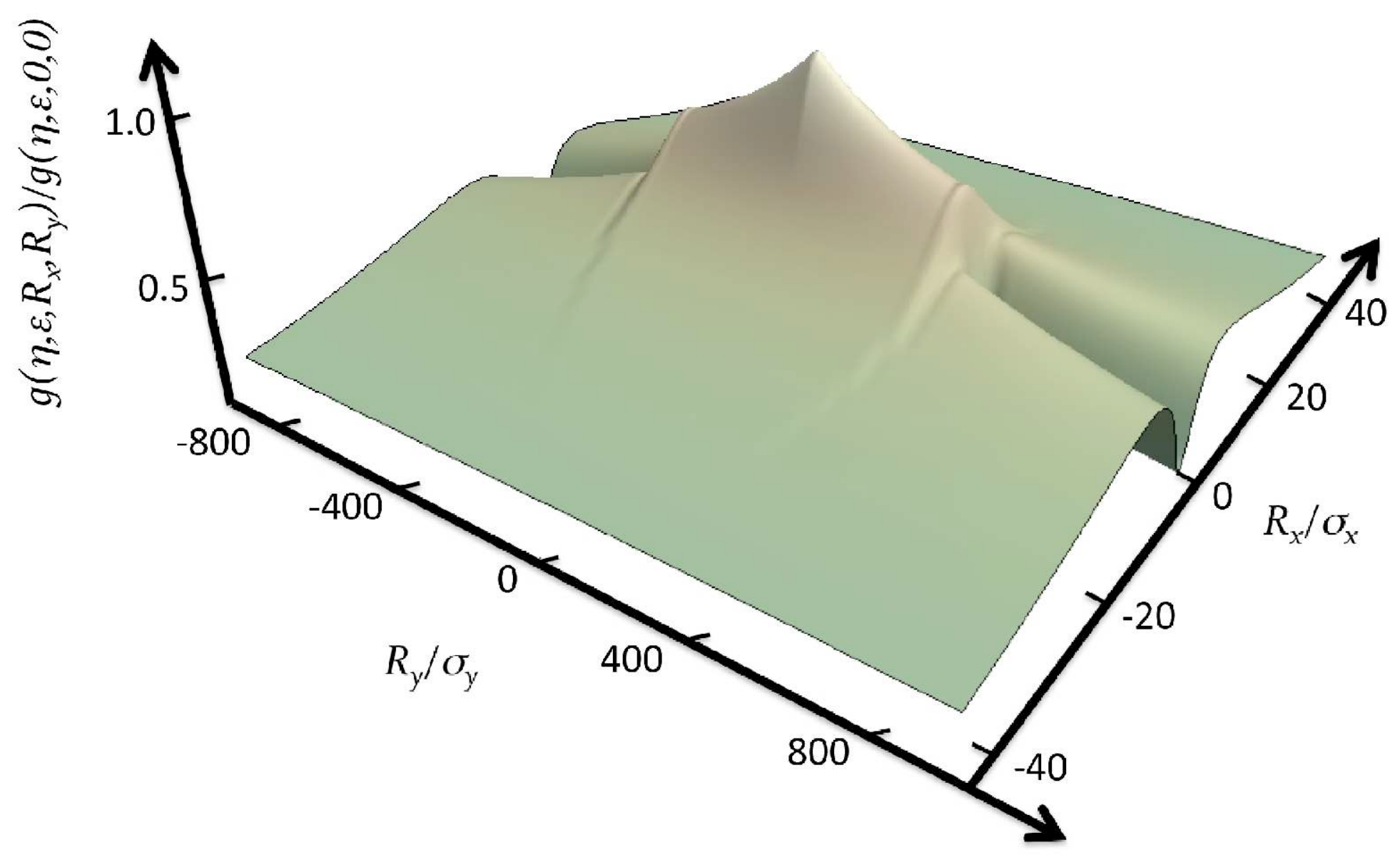} \\
{\bf (b)}
\end{center}
\end{minipage}
\caption{\label{Fig6} (Color online) For the ILC,
we plot the function $g(\eta,\varepsilon, R_x, R_y)$
as a function of the beams axis displacement $(R_x, R_y)$.
In panel (a), the region of relatively small beam axis displacement
$R_x/\sigma_x$ and $R_y/\sigma_y$ is investigated. Panel (a)
thus is a three-dimensional generalization of the curves labeled
``ILC'' in Figs.~\ref{Fig5} and~\ref{Fig6}.
Panel (b) shows the general behavior for
$g(\eta, \varepsilon, R_x, R_y)$ for ILC parameters,
covering a larger parameter range for
the beam axis displacements $R_x/\sigma_x$ and $R_y/\sigma_y$.}
\end{center}
\end{figure*}

\begin{figure*}[htb]
\begin{center}
\begin{minipage}[t]{0.49\linewidth}
\begin{center}
\includegraphics[width=1.0\linewidth]{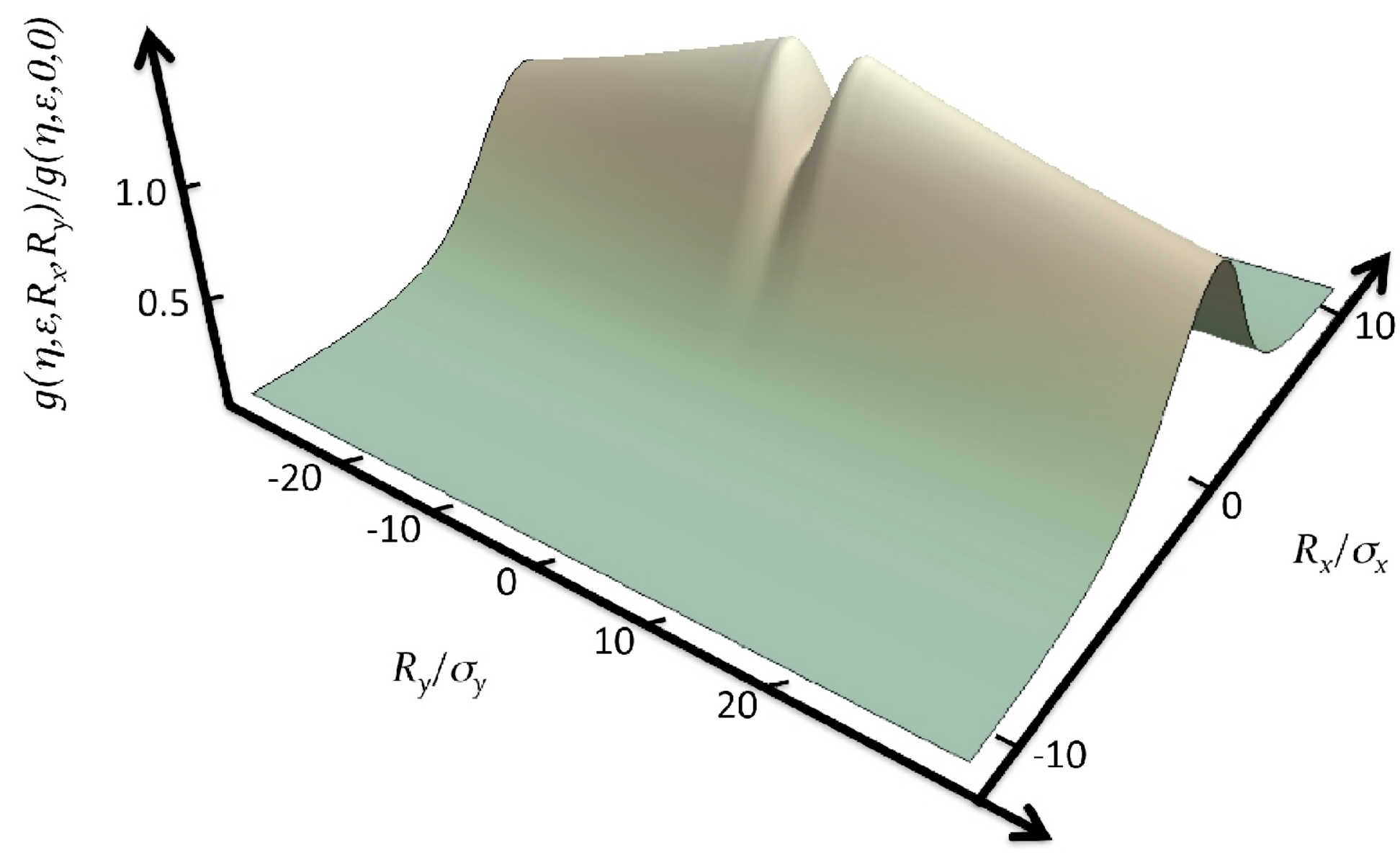} \\
{\bf (a)}
\end{center}
\end{minipage}
\begin{minipage}[t]{0.49\linewidth}
\begin{center}
\includegraphics[width=1.0\linewidth]{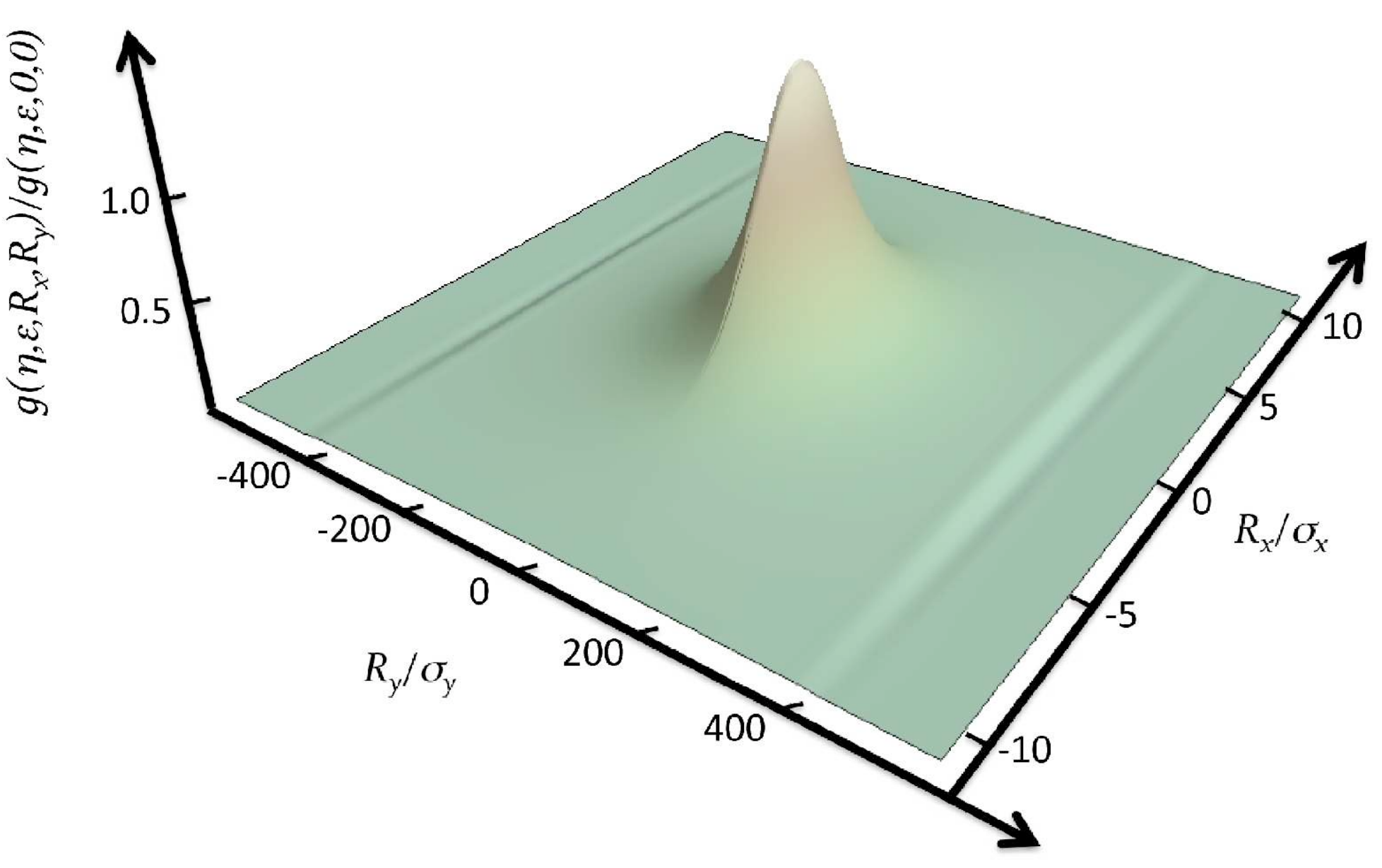} \\
{\bf (b)}
\end{center}
\end{minipage}
\caption{\label{Fig7} (Color online)
For KEKB parameters, the function $g(\eta,\varepsilon, R_x, R_y)$
is plotted as a function of $R_x$ and $R_y$.
Panel (a) shows the region of relatively small beam axis displacement
$R_x/\sigma_x$ and $R_y/\sigma_y$ and is a
three-dimensional generalization of the curves labeled
``KEKB'' in Figs.~\ref{Fig5} and~\ref{Fig6}.
The sharp ``dip'' near $x = 0$ and $y = 0$ is clearly discernible.
Panel (b) is an amplificiation of the upper one;
it shows the general behavior for
$g(\eta, \varepsilon, R_x, R_y)$ for KEKB parameters,
for larger beam axis displacements $R_x/\sigma_x$ and $R_y/\sigma_y$.}
\end{center}
\end{figure*}

%
%
\subsection{Round beams: the VEPP-2000 collider}

Now, we consider the case of round beams with $\sigma_x =
\sigma_y \equiv \sigma_\perp$ using the VEPP-2000 as an
example. If the electron bunch axis is displaced by a
distance $R$ from the positron bunch axis, the luminosity
decreases quickly for round Gaussian beams,
\begin{equation}
L(R)\;=\;L(0) \ \exp\left( -\frac{R^2}{4\sigma^2_\perp}
\right)\,.
\label{lumindecround}
\end{equation}
By contrast, the number of low-energy photons increases by
about $5$ \% at $R= 1.4\,\sigma _{\perp}$ . After that, the
rate of photons decreases, but rather slowly. Thus, at $R=
4\,\sigma _{\perp}$ the rate of photons drops by a factor
$1/2$, while the luminosity decreases by a factor $1/55$. The
corresponding curves are presented in Fig.~\ref{Fig5}. It
should be noted that curves for the
$e$-Au and $e$-$p$ mode at the
eRHIC collider have approximately the same forms as that
for the VEPP-2000 collider displayed in Fig.~\ref{Fig5}.

%
%
\subsection{Three-dimensional plots}

Round beams (VEPP-2000) imply rotational symmetry for the
functional dependence of $g(\eta, \varepsilon, R_x, R_y)$
on the beam axis displacement, and $g$ becomes
a function of $\eta$, $\varepsilon$ and $R = \sqrt{R_x^2 + R_y^2}$.
For flat beams discussed in Sec.~\ref{flatbeams}, the
situation is more complicated, and in Figs.~\ref{Fig3} and~\ref{Fig4},
the dependence of $g$ on $R_y$ for $R_x = 0$ and
on $R_x$ for $R_y=0$ is investigated.
It is instructive to generalize the treatment and to
investigate $g$ as a function of the two
variables $R_x$ and $R_y$, as done in
Figs.~\ref{Fig6} and~\ref{Fig7}, and to gradually enhance the
ranges of the beam axis displacement variables
$R_x$ and $R_y$. For flat beams
(we recall $\sigma_y \ll \sigma_x$), the electric fields
in the $y$ direction, for relatively small $y$, are
similar to the uniform fields generated by charged plates,
and this affords a rough explanation for the relatively
slow decrease of the function
$g(\eta, \varepsilon, R_x \approx 0, R_y)$ as a function of
$R_y$. However, when $R_y$ becomes so large that the
charged plates effectively shrink to a one-dimensional wire-like
structure on the scale of the beam axis displacement,
there is a rather sudden decrease of
$g(\eta, \varepsilon, R_x \approx 0, R_y)$
as a function of $R_y$ (see the lower graph in
Fig.~\ref{Fig6}).
The lower graph in Fig.~\ref{Fig7} for KEKB parameters shows an analogous
phenomenon, which however is not as clearly
discerned as for the ILC, mainly because the ILC beams
are even ``flatter'' than those for KEKB,
as is evident from the entries of Table~\ref{Table1}.
%
%
\subsection{Interpretation of the
anomalous behavior of a photon rate}
 \label{anomalous}

The peculiar existence of a maximum in the photon emission
probability at non-vanishing beam axis displacement,which
we found for both round and flat beams, can be explained as
follows. Let us discuss flat beams with $\sigma_y \ll
\sigma_x$, for definiteness. At $R_{x}=R_{y}=0$, a
considerable portion of the electrons moves in the region
of small impact parameters where the electric and magnetic
fields of the positron bunch are small. For the vertical
displacement $R_{y}$ in the range of $\sigma _{y}^2 \ll
R_{y}^2 \ll \sigma _{x}^2$, the electrons fly through a
stronger electromagnetic field of the positron bunch, and
therefore, the number of emitted photons increases. At
larger $R_y$ (for $\sigma _x^2 \ll R_y^2 \ll \sigma_z^2$),
the fields of the positron bunch are $|{\bm E}| \approx
|{\bm B}| \propto\; |1/R_y|$, which leads to $\dd N_\gamma
\propto \; 1/R_y^2$. In that region, the number of emitted
photons decreases, but only polynomially as compared to the
exponential drop of the luminosity.
By contrast, when shifting the electron bunch axis for flat
beams in the horizontal direction, the positron bunch
fields as seen by the electrons
immediately become weaker, and the photon rate slowly
decreases.

%
%
\section{Summary}
\label{sec4}

We have considered the emission of radiation in the
collisions of an electron bunch with a dense positron (or
ion) bunch at modern colliders, with relevant parameters
given in Table~\ref{Table1}. In this case, the electron
deflection angle is of the order of $\eta/\gamma_e$ with
the parameter $\eta$ being not small. For the ILC collider,
we have
$\eta=87 \gg 1$, and there are three different regimes of
photon emission determined by two critical photon
energies $E_{\rm c}^{\rm BS}$ and $E_{\rm c}$ given in
Eqs.~(\ref{EcBC}) and (\ref{Ec}), respectively. For
large photon energies, $E_\gamma \gg E_{\rm c}^{\rm BS}$,
the emission of photons is an incoherent process described
by the corresponding Feynman diagrams for binary collisions.
For smaller photon energies, coherent phenomena become
important, and the electrons interact with the collective
electromagnetic field of the incoming bunch.

In the photon energy range $E_{\rm c} \ll E_\gamma \lesssim
E_{\rm c}^{\rm BS}$, the emission of photons is described
by the beamstrahlung approximation similar to the usual
synchrotron radiation. In this case, the electron coherence
length is small as compared to the length of the incoming
bunch. For even smaller photon energies, $E_\gamma < E_{\rm
c}$, the coherence length over which the electron has to
travel in order to accommodate for the formation length,
becomes comparable with the bunch length $\sigma_z$ or even
larger, and therefore, one has to use yet a different
approximation based on classical currents, as detailed in
the current paper.

The classical-current process dominates over
the beamstrahlung emission in
the range of small photon energies, as evident from
Fig.~\ref{Fig2}, and there is a characteristic maximum in
its dependence on the beam axis displacement, as evident
from Figs.~\ref{Fig3} and ~\ref{Fig5}. An explanation for
the latter observation is provided in Sec.~\ref{anomalous}.
The existence of the maximum could even be used
for a fast ``online''
control of the beam axis displacement.

Let us conclude this paper by listing a few concrete
numbers. For the ILC collider, the critical energy, below
which the large-coherence-length approach should be used,
lies at $E_{\rm c} \approx 83$~keV, and the number of
photons emitted per $N_e$ electrons per bunch crossing is
\begin{equation}
{\dd N_\gamma}\approx 5.5\,\alpha\,{N_e} \frac{\dd
E_\gamma}{E_\gamma} \,.
\end{equation}
In general, our study highlights the need for an accurate
understanding of electromagnetic processes at the discussed
colliders (Tables~\ref{Table1} and~\ref{Table2}); these
which may be sources of a number of problems
related to energy losses and background. 

In principle, the low-energy photons discussed here could also be used in order
to directly monitor the bunch sizes at their interaction points. At least, one
such possibility is pointed out in an experiment which has already been
successfully performed at the SLC collider~\cite{BoEtAl1989} with an electron
energy of about 50~GeV.  In this experiment, a transverse size of the electron
bunch of the order of 10~$\mu$m was measured just by detecting photons of
relatively small energy.

%
%
\section{Acknowledgments}

We are very grateful to A.~Bondar, I.~Ginzburg, N.~Shulga  and
V.~Telnov and the late G.~Soff for useful discussions. V.G.S.
acknowledges warm hospitality of the Institute of Theoretical
Physics of Heidelberg University and support by the Gesellschaft
f\"{u}r Schwerionenforschung (GSI Darmstadt) under contract
HD--JENT. This work is partially supported by the Russian Foundation
for Basic Research (code 09-02-00263) and by the Fond of Russian
Scientific Schools (code 1027.2008.2).

\appendix

%
%
\section{Approximation of classical currents}
\label{appa}

We attempt to reproduce the main result of
Ref.~\cite{ShTy2003} by a different (shorter) proof. The case
under consideration in Ref.~\cite{ShTy2003} is the head-on
collision of one electron with the ion bunch. In the
approximation of classical currents, one can calculate the
number of photons produced by this electron just by using the
classical expression for the radiated energy $\dd {\cal E}$
divided by the photon energy $E_\gamma$ and relate this to
the probability of emitting a photon $\dd w$
(see~\cite{BeLiPi1982vol4}, \S~98),
\begin{equation}
\dd w = \frac{\dd{\cal E}}{E_\gamma}
      = \alpha\,G(\xi)\,
        \frac{\dd E_\gamma}{E_\gamma}\,,
\end{equation}
where the function $G(\xi)$ given in Eq.~(\ref{Fxi}) has
the following asymptotic behaviour,
\begin{equation}
G(\xi) = \left\{\begin{array}{ll} {\displaystyle
\frac{8}{3\pi}} \, \left(\xi^2 -{\displaystyle\frac 35}
\xi^4\right)\; & \mbox{ at } \;\; \xi^2 \ll 1\,,
\\[4ex]
{\displaystyle \frac{2}{\pi}} \,
\left[\ln\left(4\xi^2\right)-1 \right] \; &
\mbox{ at } \;\; \xi^2 \gg 1\,.
\end{array}\right.
\end{equation}
The dimensionless parameter
\begin{equation}
\xi = \frac{|{\bm p}|}{m_ec} \, \sin\left( \frac{\theta_e}{2} \right) =
\frac{|{\bm p}_\perp|}{2m_ec}
\end{equation}
is expressed via the transverse momentum ${\bm p}_\perp$
acquired by electron during the collision. If an electron
with the impact parameter $\mbox{\boldmath $\varrho$}$
moves in the field of one nucleus with the charge $Ze$, it
gets the transverse momentum (see~\cite{LaLi1960vol2}, problem 2 in
\S~39)
\begin{equation}
{\bm p}_\perp = - \frac{2Ze^2}{c}\,
\frac{\mbox{\boldmath $\varrho$}}{\varrho^2}\,.
\label{A.p-perp-eZ}
\end{equation}
Therefore, the parameter $\xi$ for the collision of one
electron with the ion bunch is
\begin{equation}
\mbox{\boldmath $\xi$}(\mbox{\boldmath $\varrho$}) =
\frac{{\bm p}_\perp}{2m_ec} =
-Z\,r_e\, \int
\frac{\mbox{\boldmath $\varrho$} - {\mbox{\boldmath $\varrho$}}'}%
{({\mbox{\boldmath $\varrho$}} - \mbox{\boldmath
$\varrho$}')^2} \, n_p({\mbox{\boldmath $\varrho$}}')\,
\dd^2 \varrho'\,, \label{A.xi}
\end{equation}
where $n_p({\mbox{\boldmath $\varrho$}})$ is the transverse
density of the ion bunch. Finally, we would like to mention
that in the paper~\cite{ShTy2003}, there are misprints in
formulae corresponding to those given here in
Eqs.~(\ref{A.p-perp-eZ}) and~(\ref{A.xi}).

%
%
\section{Case of Gaussian beams}
\label{appb}
%
%
\subsection{General case}

We start with the calculation of $\xi_x(\mbox{\boldmath
$\varrho$})$ from Eq.~(\ref{A.xi}), using the following
integral representation:
\begin{align}
\frac{\varrho_x - \varrho_x'}%
{(\mbox{\boldmath $\varrho$}-\mbox{\boldmath $\varrho$}')^2}
=& \; -\frac12 \,\int_0^{1} \, \frac{\dd u}{u(1-u)}
\\[2ex]
& \; \times \frac{\partial}{\partial \varrho_x}\,
\exp{\left[-\frac{u}{1-u}\,\frac{(\mbox{\boldmath
$\varrho$}-\mbox{\boldmath
$\varrho$}')^2}{2\sigma_x^2}\right]}\,.
\nonumber
\end{align}
Under the above transformation, the integrals over
$\varrho'_x$ and  $\varrho'_y$ for the Gaussian beam
(\ref{9}) become of the Gaussian type:
\begin{equation}
\int_{-\infty}^{\infty} {\rm e}^{-at^2+2bt}dt=
\sqrt{\frac{\pi}{a}}\,{\rm e}^{b^2/a}\,.
\label{Gauss}
\end{equation}
Indeed, if we introduce new, scaled variables,
\begin{equation}
x=\frac{\varrho_x}{\sqrt{2} \sigma_x}\,,\;
x'=\frac{\varrho'_x}{\sqrt{2} \sigma_x}\,,\;
y=\frac{\varrho_y}{\sqrt{2} \sigma_y}\,,\;
y'=\frac{\varrho'_y}{\sqrt{2} \sigma_y}\,,
\end{equation}
then the quantity $\xi_x(\mbox{\boldmath $\varrho$})$ takes the
form:
\begin{align}
& \xi_x(x,y) = \frac{N_pZr_e}{2\sqrt{2}\, \pi \sigma_x}
\nonumber\\[2ex]
& \; \times \int_0^1\frac{\dd u}{u\,(1-u)}\,
\frac{\partial}{\partial x}\,\int_{-\infty}^{\infty}
dx'\,\int_{-\infty}^{\infty} dy' \;{\rm e}^{-\Phi}\,,
\end{align}
where
\begin{eqnarray}
\Phi&=& a_x x'^2-2b_xx' +c_x +a_y y'^2-2b_yy' +c_y\,,
\nonumber
\\
a_x&=&\frac{1}{1-u},\;\;
b_x=\frac{ux}{1-u}\,,\;
c_x=\frac{ux^2}{1-u}\,,
\\
a_y&=&\frac{1-(1-\varepsilon^2)u}{1-u},\;\;
b_y=\frac{u\varepsilon^2y}{1-u}\,,\;
c_y=\frac{u\varepsilon^2y^2}{1-u}\,.
\nonumber
\end{eqnarray}
and the beam aspect ratio $\varepsilon$ is given in
Eq.~(\ref{ar}). Using Eq.~(\ref{Gauss}) we immediately
obtain
\begin{equation}
\xi_x(x,y)= -\frac{\eta\,(1+\varepsilon)}{\sqrt{2}}\,x\,
\int_0^1 \ee^{-A}\,
\frac{\dd u}{\sqrt{1-(1-\varepsilon^2)u}}\,,
\label{xix}
\end{equation}
where
\begin{equation}
A= u \, x^2+
\frac{u\varepsilon^2}{1-(1-\varepsilon^2)u}\,y^2\,.
\end{equation}
and
\begin{equation}
\eta=\frac{N_p Zr_e}{\sigma_x+\sigma_y}\,,\;\;
\varepsilon = \frac{\sigma_y}{\sigma_x}\,.
\end{equation}
The quantity $\xi_y(\mbox{\boldmath $\varrho$})$ can be
obtain from this expression by simple replacements,
\begin{equation}
\varrho_x \leftrightarrow \varrho_y \,,
\qquad
\sigma_x \leftrightarrow \sigma_y \,,
\end{equation}
or $x\leftrightarrow y,\;\varepsilon\to 1/\varepsilon$ and
reads
\begin{align}
\xi_y(x,y) =& \;
-\frac{\eta\,(1+\varepsilon)}{\sqrt{2}}\,y\, \int_0^1
\ee^{-B}\, \frac{\dd v}{\sqrt{1-(1-\varepsilon^2)v}}\,,
\nonumber\\[2ex]
B=& \; (1-v) y^2 + \frac{1-v}{1 - (1-\varepsilon^2)v}\,x^2
\label{xiy}
\end{align}
(with the substitution $u=1-v$).

Now, using the expressions (\ref{xix}), (\ref{xiy}) and
Eqs.~(\ref{dNgamma2}), (\ref{dis}), we obtain for
identical Gaussian beams the important function
\begin{equation}
g(\eta, \varepsilon, R_x, R_y) = \frac{1}{\pi}
\int\limits_{-\infty}^{\infty} \dd x
\int\limits_{-\infty}^{\infty} \dd y \,
\frac{G(\xi(x,y))}{\ee^{(x-X)^2+(y-Y)^2}} \,,
\end{equation}
where
\begin{equation}
X=\frac{R_x}{\sqrt{2}\sigma_x}\,,\;\;
Y=\frac{R_y}{\sqrt{2}\sigma_y}\,.
\end{equation}
This equation is used in the current study for numerical
calculations.

%
%
\subsection{Round beams}

For round identical beams, we have $\sigma_x = \sigma_y
\equiv \sigma_\perp$, and the parameter $\eta$ becomes
\begin{equation}
\label{defetaround} \eta= \frac{Z r_e N_p}{2\sigma_\perp} \,.
\end{equation}
Some formulae are thus simplified. The integral in
Eq.~(\ref{A.xi}) can be easily calculated,
\begin{eqnarray}
&&\xi =\sqrt{\xi_x^2+\xi_y^2}=
\frac{\sqrt{2}\,\eta}{\sqrt{x^2+y^2}}\, \left( 1-{\rm
e}^{-x^2-y^2}\right)\,,
 \nonumber\\
&&\sqrt{x^2+y^2}= \frac{\varrho}{\sqrt{2}\sigma_\perp}\,,
\end{eqnarray}
and the function $g$ becomes equal to
\begin{align}
g =& \; \int_0^{\infty} G(\xi(\varrho)) \;
\exp{\left(-\frac{R^2+\varrho^2}{2\sigma^2_{\perp}}\right)}\,
\nonumber\\[2ex]
& \; \times I_0\left( \frac{\varrho \, R}{\sigma^2_{\perp}}\right) \;
\frac{ \varrho\,\dd \varrho}{\sigma^2_{\perp}}\,,
\end{align}
where $I_0(x)$ is the modified Bessel function of the first
kind. At $R/\sigma_\perp \gg \eta$, this function is
\begin{equation}
g= \frac{32}{3\pi} \,\eta^2 \,
\left(\frac{\sigma_\perp}{R}\right)^2
 \label{smalleta}
\end{equation}
in accordance with the CBS result of
Refs.~\cite{GiKoPoSe1992prl,GiKoPoSe1992plb,GiKoPoSe1993,GiKoPoSe1992yf1,GiKoPoSe1992yf2}.

The behavior of $g$ as a function of $\eta$ at
$R=0$ is the following. At small values of $\eta$, the
function $g(\eta)\equiv g(\eta,\varepsilon=1, R_x=0,R_y=0)$
is
\begin{equation}
g(\eta) = a \, \eta^2 - b \, \eta^4\;, \qquad
\eta^2 \ll 1\,,
\label{s}
\end{equation}
where
\begin{align}
a =& \; \frac{16}{3\pi}\, \ln\left(\frac43 \right) =
0.4884\,, \\
b =& \; \frac{32}{5\pi} \, \left(5\,\ln{5} + 18\,\ln{3} -
40 \, \ln{2}\right) = 0.1962\,.
\nonumber
\end{align}
The coefficient $a$ coincides with that in CBS, and the
ratio $a/b=0.4018$ indicates the range of applicability of
the CBS approach. At large values of~$\eta$,
\begin{subequations}
\begin{align}
g(\eta) =& \;  \frac{2}{\pi}\, \left(\ln{\eta^2} -c\right) \,, \qquad
\eta^2 \gg 1\,,
\label{largeeta}
\\[1ex]
c =& \; 3 - \gamma_\mathrm{E} - 3 \, \ln{2} = 0.3433,\,
\end{align}
\end{subequations}
and $\gamma_\mathrm{E} = 0.5772\ldots$ is Euler's constant.
Equations~(\ref{s}) and (\ref{largeeta}) give us
approximations with an accuracy better than $5\;\%$
for small and large $\eta$, i.e.~outside the region $\eta=0.7\div 4$,
for round beams at zero beam axis displacement.

\end{document}